\documentclass{article}
\pdfoutput=1
\usepackage{amsmath}
\usepackage{amsfonts}
\usepackage{amssymb}
\usepackage{graphicx}
\begin{document}

\title{Entropic Inference: some pitfalls\\and paradoxes we can avoid\thanks{Presented at MaxEnt 2012, The 32nd
International Workshop on Bayesian Inference and Maximum Entropy Methods in
Science and Engineering, (July 15--20, 2012, Garching, Germany). } }
\author{Ariel Caticha\\{\small Department of Physics, University at Albany-SUNY, }\\{\small Albany, NY 12222, USA.}}
\date{}
\maketitle

\begin{abstract}
The method of maximum entropy has been very successful but there are cases where it has either failed or led to paradoxes that have cast doubt on its general legitimacy. My more optimistic assessment is that such failures and paradoxes provide us with valuable learning opportunities to sharpen our skills in the proper way to deploy entropic methods. 

The central theme of this paper revolves around the different ways in which constraints are used to capture the information that is relevant to a problem. This leads us to focus on four epistemically different types of constraints. I propose that the failure to recognize the distinctions between them is a prime source of errors.

I explicitly discuss two examples. One concerns the dangers involved in
replacing expected values with sample averages. The other revolves around misunderstanding ignorance. I discuss the Friedman-Shimony paradox as it is manifested in the three-sided die problem and also in its original thermodynamic formulation.
\end{abstract}

\section{Introduction}

Our subject is entropic inference. The method of maximum entropy (whether in
its original MaxEnt version or its generalization ME, the method for
updating probabilities) has been successful in many applications but there
are cases where it has either failed or led to paradoxes. It has been
suggested that these are symptoms of irreparable flaws. I disagree. My
assessment is considerably more optimistic: the paradoxes provide us with
valuable opportunities for learning how to use the method and equally
valuable warnings about pitfalls we can and should avoid.

First, some background.\footnote{%
I make no attempt to provide a review of the literature on entropic
inference. The following list, which ref{}lects only some contributions that
are directly related to the particular approach described in this tutorial,
is incomplete but might nevertheless be useful: Jaynes \cite{Jaynes 2003},
Shore and Johnson \cite{Shore Johnson 1980}, Williams \cite{Williams 1980},
Skilling \cite{Skilling 1988}, Rodr\'{\i}guez \cite{Rodriguez 1991}, Giffin
and Caticha \cite{Caticha Giffin 2006}-\cite{Caticha 2012}.} The objective
of the method of maximum entropy (ME) is to update from a prior distribution 
$q$ to a posterior distribution when information is given that the posterior 
$P$ is constrained to belong to a certain family of distributions: $P\in 
\mathcal{C}=\{p\}$. The selected posterior $P$ is that which maximizes the
(relative) entropy 
\begin{equation}
S[p,q]=-\int dx\,p(x)\log \frac{p(x)}{q(x)}~  \label{S[pq]}
\end{equation}%
subject to the constraint $\mathcal{C}$. Justifying the method revolves to a
large extent around justifying the particular choice of the functional $%
S[p,q]$. The criteria involved in designing the functional $S[p,q]$ are
purely pragmatic:

\noindent \textbf{(1)} We seek a method of universal applicability. It is
conceivable that different situations could require different induction
methods but \emph{what we want is a general-purpose method that captures
what all those other problem-specific methods have in common.}\footnote{%
This approach to entropic inference includes Jaynes' MaxEnt method and
Bayesian inference as special cases. The MaxEnt method is recovered when the
prior $q$ reflects an underlying measure or (up to a normalization factor) a
uniform distribution. Bayesian inference is recovered when the goal is to
infer parameters $\theta $ on the basis of information about data $x$ and
the relation between $x$ and $\theta $ as given by a known likelihood
function $q(x|\theta )$. \cite{Williams 1980}\cite{Caticha Giffin 2006}}

\noindent \textbf{(2)} We want a parsimonious method that recognizes the
value of information. What has been laboriously learned in the past should
not be disregarded unless rendered obsolete by new information. \emph{Priors
matter: rational beliefs should be updated but only to the minimal extent
demanded by the new information.}

\noindent \textbf{(3)} The method must be useful in practice. In particular,
in order to do science we must be able to understand parts of the universe
without having to understand the universe as a whole. This implies that the
notion of statistical independence must play a central and privileged role.
This idea -- that some things can be neglected, that not everything matters
-- is implemented by imposing a criterion that tells us how to handle
independent systems. The design criterion we adopt is quite natural: \emph{%
Whenever two systems are a priori believed to be independent and we receive
information about one it should not matter if the other is included in the
analysis or not.}\footnote{%
This amounts to requiring that independence be preserved unless information
to the contrary is explicitly introduced which is in accordance with the
principle of parsimony stated in \textbf{(2)}.}

\noindent The subtleties of how these criteria are implemented by imposing
locality, coordinate invariance, and independence and the proofs showing
that they lead to the entropy functional (\ref{S[pq]}) are discussed in \cite%
{Caticha 2012}. A noteworthy feature is that it is not necessary to provide
an interpretation for $S[p,q]$ be it in terms of heat, or disorder, or
amounts of information. Entropy is a tool for updating that requires no
further interpretation.

The task of entropic inference -- to update rational beliefs when
information becomes available -- immediately points both to the question
\textquotedblleft What is information?\textquotedblright\ and to its answer.
A fully Bayesian information theory demands a close relation between
information and the beliefs of an ideally rational agent and, accordingly,
the answer is: \emph{information is that which affects rational beliefs, and
thus, constraints are information. }Which leads us to the main topic of this
paper:\ how to deploy constraints that correctly capture the information
that is relevant to a problem. In the next section I argue that we ought to
distinguish four epistemically different cases and that the failure to
recognize the distinctions among them is a prime source of mistakes and
paradoxes.

Over the years a number of objections have been raised against the method of
maximum entropy. I\ believe some of these objections were quite legitimate
at the time they were raised. They uncovered conceptual pitfalls with the
old MaxEnt as it was understood at the time. I also believe that in the
intervening decades our understanding of entropic inference has evolved to
the point that all these concerns can now be addressed satisfactorily. I
explicitly discuss two examples of such mistakes. One concerns the dangers
of replacing expected values with sample averages (a good discussion of this
so-called \textquotedblleft constraint rule\textquotedblright\ is \emph{e.g.}
\cite{Uffink 1996}). The other revolves around misunderstanding ignorance.%
\footnote{%
We use `ignorance' as a technical term to denote lack of knowledge or lack
of information. The term `uncertainty' might also be apropriate except that
through its heavy use in so many other contexts it has acquired other
connotations that could be misleading. We do not mean to suggest that a
situation of ignorance is in any way morally reprehensible.} These are
objections of the type raised by the Friedman-Shimony paradox \cite{Friedman
Shimony 1971}\cite{Shimony 1985} as it is manifested in the three-sided die
problem \cite{Uffink 1996}\cite{Seidenfeld 1986} and also in its original
thermodynamic formulation.\footnote{%
For additional references on the controversy around the Friedman-Shimony
paradox see \cite{Uffink 1996}. For yet another paradox that can be handled
with the arguments we deploy here see \cite{Seidenfeld 1986}\cite{van
Fraasen 1981}. Other objections raised by these authors, such as the
compatibility of Bayesian and entropic methods, have been fully addressed
elsewhere.\cite{Caticha Giffin 2006}\cite{Caticha 2012}}

\section{On constraints and relevant information}

To fix ideas consider the standard MaxEnt problem: to assign the probability
of a discrete variable $i$ assuming a uniform underlying measure $q_{i}=$ 
const and information in the form of a single linear constraint, $%
\langle f\rangle =F$. MaxEnt requires us to maximize the Shannon entropy $%
S[p]=-\textstyle\sum_{i} p_{i}\log p_{i}$ subject to $\langle f\rangle =F$
and $\textstyle\sum_{i} p_{i}=1$ which yields $p(i|\lambda )\propto e^{-\lambda f_{i}}$
for an appropriately chosen $\lambda $. 

For example, the canonical distribution that describes the state of
thermodynamic equilibrium is obtained maximizing $S[p]$ subject to a
constraint on the expected energy $\langle \varepsilon \rangle =E$. This
yields the Boltzmann distribution, $p(i|\beta )\propto e^{-\beta \varepsilon
_{i}}$, where $\beta =1/T$ is the inverse temperature. The questions that
concern us here are: How do we decide which is the right constraint function 
$f$ to choose? How do we decide the numerical value $F$ of its expectation?
When can we expect the inferences to be reliable?

When using the MaxEnt method to obtain, say, the Boltzmann distribution it
has been common to adopt the following language:

\begin{description}
\item[\textbf{\qquad }] We seek the probability distribution that codifies
the information we actually have (\emph{e.g.}, the expected energy) and is
maximally unbiased (\emph{i.e.} maximally ignorant or maximum entropy) about
all the other information we do not possess.\ 
\end{description}

\noindent This justification has stirred considerable controversy. Some of
the objections that have been raised are the following:

\begin{description}
\item[\textbf{O1}] Objective reality is independent of our subjective
knowledge about it. The observed spectrum of black body radiation is what it
is independently of whatever information happens to be available to us.

\item[\textbf{O2}] In most realistic situations the expected value of the
energy is not a quantity we happen to know. How, then, can we justify using
it as information we actually have?

\item[\textbf{O3}] Even when the expected values of some quantities happen
to be known, there is no guarantee that the resulting inferences will be any
good at all.\ How can we justify the success of thermodynamics?
\end{description}

\noindent These objections deserve our consideration.

The issue raised by \textbf{O1} concerns the very essence of what physical
theories are meant to be. Let us grant for the sake of argument that there
is such a thing as an external reality, that real phenomena are what they
are independently of our thoughts about them. Then the issue raised by 
\textbf{O1} is whether the purpose of our theories is to provide \emph{%
models that faithfully mirror this external reality} or whether the
connection to reality is considerably more indirect and the \emph{models are
pragmatic tools for manipulating information about reality} for the purposes
of prediction, control, explanation, etc. In the former case theories mirror
reality and there is no logical room for subjectivity. In the latter case
theories deal with \emph{our} information about reality and some subjective
elements are inevitable. The evidence in favor of the latter alternative is
already considerable. It includes the successful derivation and a host of
insights into statistical mechanics and thermodynamics achieved by Jaynes'
MaxEnt. It also includes the more recent entropic/Bayesian derivations of
both quantum and classical mechanics. 

The objection \textbf{O1} originates in a failure to recognize that while
those epistemic judgments that must be made when assigning probabilities are
inevitably subjective, they can nevertheless still be objectively right or
wrong depending on whether they achieve empirical success or not. Thus, we
can evade \textbf{O1} by refusing to wear a straightjacket that forces us
into a strict subjective/objective dichotomy. Subjective judgments do not
preclude inferences that are objectively correct or incorrect.

To address objections \textbf{O2} and \textbf{O3} it is useful to
distinguish four epistemically different types of constraints:

\begin{description}
\item[(A)] We know that the expected value of the function $f$ captures
information that happens to be relevant to the particular problem at hand
and we also know its numerical value, $\langle f\rangle =F$.
\end{description}

\noindent The ideal situation is one in which the set of available
constraints of type \textbf{A} is complete in the sense that all the
information that is necessary to obtain reliable answers to the questions
that interest us is available. Only then are we guaranteed reliable
predictions.\footnote{%
Our goal here has been merely to describe the epistemically ideal situation
one would like to achieve. The important question of how to assess whether a
particular set of constraints is relevant and complete for any specific
issue at hand will not be addressed here except to point out that the
criteria of success are purely pragmatic. More specifically, Jaynes has
suggested that the appropriate criterion is \emph{reproducibility}:
\textquotedblleft If any macrophenomenon is found to be reproducible, then
it follows that all microscopic details that were not reproduced, must be
irrelevant for understanding and predicting it. In particular, all
circumstances that were not under the experimenter's control are very likely
not to be reproduced, and therefore are very likely not to be
relevant.\textquotedblright\ \cite{Jaynes 1985} (See also Section 5.8 of 
\cite{Caticha 2012}.)} Both requirements of relevance and completeness are
crucial: an incomplete set of type \textbf{A} constraints has predictive
value in that it leads to the best predictions that one can achieve under
the circumstances but there is no guarantee that the predictions will be any
good. Thus we see that, properly understood, objection \textbf{O3} is not a
flaw of the entropic method; it is a legitimate warning that reasoning with
incomplete information is a risky business.

Note that a particular piece of evidence can be relevant and complete for
some questions but not for others. For example, the expected energy $\langle
\varepsilon \rangle =E$ is both relevant and complete for the question
\textquotedblleft Will system 1 be in thermal equilibrium with another
system 2?\textquotedblright\ or alternatively, \textquotedblleft What is the
temperature of system 1?\textquotedblright\ But the same expected energy is
far from complete for the vast majority of other possible questions such as,
for example, \textquotedblleft Where can we expect to find molecule \#23 in
this sample of ideal gas?\textquotedblright\ 

\begin{description}
\item[(B)] We know that $\langle f\rangle $ captures information that
happens to be relevant to the problem at hand but its actual numerical value 
$F$ is not known.
\end{description}

\noindent This is the most common situation in physics. The answer to
objection \textbf{O2} hinges on the observation that whether the value of
the expected energy $E$ is known or not, it is nevertheless still true that
maximizing entropy subject to the energy constraint $\langle \varepsilon
\rangle =E$ leads to the objectively correct \emph{family} of distributions
that describe thermal equilibrium (including, for example, the observed
black-body spectral distribution). Thus, the justification behind imposing a
constraint on the expected energy is not that the quantity $E$ happens to be
known -- because of the brute fact that its value is never actually known --
but rather that it is a quantity that \emph{should }be\emph{\ }known. Even
when the actual numerical value $E$ is unknown, the epistemic situation
described in case \textbf{B} is one in which we know that the expected
energy $\langle \varepsilon \rangle $ is the \emph{relevant} information
without which no successful predictions are possible.

The question of how a particular $f$ is singled out as relevant has to be
tackled on a case by case basis. In \cite{Caticha 2012} we discuss the
problem of thermal equilibrium and show that the relevant quantity is indeed
the expected energy $\langle \varepsilon \rangle $ and not some other
conserved quantity such as $\langle \varepsilon ^{2}\rangle $ or $\langle
f(\varepsilon )\rangle $.

Type \textbf{B} information is processed by allowing MaxEnt to proceed with
the numerical value of $\langle \varepsilon \rangle =E$ handled as a free
parameter. This leads us to the correct \emph{family }of distributions $%
p(i|\beta )\propto e^{-\beta \varepsilon _{i}}$ containing the multiplier $%
\beta $ as a free parameter. The actual value of the parameter $\beta $ is
at this point unknown and the standard approach is to seek additional
information and infer $\beta $ either by a direct measurement using a
thermometer, or to infer it indirectly by Bayesian parameter estimation from
other empirical data. The additional information has the net effect of
transforming the type \textbf{B} constraint into a type \textbf{A}.

\begin{description}
\item[(C)] There is nothing special about the function $f$ except that we
happen to know its expected value, $\langle f\rangle =F$. In particular, we
do not know whether information about $\langle f\rangle $ is complete or
whether it is at all relevant to the problem at hand.
\end{description}

\noindent We do know something and this information, although limited, has
some predictive value because it serves to constrain our attention to the
subset of probability distributions that agree with it. Maximizing entropy
subject to such a constraint will yield predictions that are the best
possible under the circumstances but since we do not know that $f$ captures
information that is relevant and complete there is absolutely no guarantee
that the predictions will be any good. Induction is risky and objection 
\textbf{O3} is a healthy reminder.

\begin{description}
\item[(D)] We know neither that $\langle f\rangle $ captures relevant
information nor do we know its numerical value $F$.
\end{description}

\noindent This is an epistemic situation that reflects complete ignorance.
Case \textbf{D} applies to any arbitrary function $f$ and therefore it
applies equally to all functions. Since no specific $f$ is singled out a
type \textbf{D} constraint provides no information at all and the correct
procedure is to maximize $S[p]$ subject to the single constraint of
normalization. The result is as it should be: extreme ignorance is described
by a uniform distribution.

What distinguishes type \textbf{C} from \textbf{D} is that in \textbf{C} the
value of $F$ is actually known. This fact singles out a specific variable $f$
and justifies using $\langle f\rangle =F$ as a constraint. What
distinguishes \textbf{D} from \textbf{B} is that in \textbf{B} there is
actual knowledge that singles out the variable $f$ as being \emph{relevant}.

Objection \textbf{O2} arises from a failure to distinguish constraints of
type \textbf{B} from those of type \textbf{C} and \textbf{D}.

\noindent \textbf{Summary:} Between one extreme of ignorance (type \textbf{D}%
, we know neither which variables are relevant nor their expected values),
and the other extreme of useful knowledge (a complete set of type \textbf{A}
constraints in which we know all the relevant variables that need to be
included in the analysis and we also know their expected values), there are%
\emph{\ intermediate states of knowledge} (involving constraints of types 
\textbf{B} and \textbf{C}) and these constitute the rule rather than the
exception. (It is, of course, also possible to encounter situations that mix
constraints of different types.) Type \textbf{B} is the more common and
important situation in which a relevant variable has been correctly
identified even though its actual expected value might be unknown. The
situation described as type \textbf{C} is less common because information
about expected values is not usually available. (What might be easily
available is information in the form of sample averages which is not in
general quite the same thing --- see the next section.) Type \textbf{D}
constraints carry no information at all and should be ignored.

The considerations above also apply to the problem of entropic updating from
a generic prior $q$ to a posterior by maximizing $S[p,q]$ subject to
information in the form of constraints.

\section{Sample averages are not expected values}

Let us return to the question \textquotedblleft If constraints refer to the
expectation of certain variables, how do we decide their numerical
magnitudes?\textquotedblright\ and explore some pitfalls. Here is a common
temptation: the numerical values of expectations are seldom known and it is
tempting to follow the \textquotedblleft constraint rule\textquotedblright\
and replace expected values by sample averages because it is the latter that
are directly available from experiment. But the two are not the same: \emph{%
Sample averages are experimental data. Expected values are not experimental
data.} 

For very large samples such a replacement can be justified by the law of
large numbers --- there is a high probability that sample averages will
approximate the expected values. However, for small samples using one as an
approximation for the other can lead to incorrect inferences. It is
important to realize that these incorrect inferences do not represent an
intrinsic flaw of the entropic method; they are merely an indication of how
the method should not to be used.

\noindent \textbf{Example --- just data:}

Here is a variation on the same theme. Suppose data $D=(x_{1},x_{2}\ldots
x_{n})$ have been collected. How do we process such information? Suppose we
do not have a likelihood function so Bayes rule is not an option. We might
be tempted to maximize $S[p,q]$ subject to a constraint $\langle x\rangle
=C_{1}$ where $C_{1}$ is unknown and then try to estimate $C_{1}$ as a
sample average. This is a dangerous move. The reason is that \emph{in the
absence of additional information} we know neither that $x$ constitutes
relevant information nor do we know its expected value $C_{1}$ and therefore
this is what we identified above as a type \textbf{D} constraint --- no
information at all.

The mistake becomes apparent when we realize that if we know the data $%
(x_{1},\ldots )$ then we also know their squares $(x_{1}^{2},\ldots )$ and
their cubes and also any arbitrary function of them $(f(x_{1}),\ldots )$.
And we also know the corresponding sample averages. Which of these should we
use as an expected value constraint? Or should we use all of them? The
answer is that the entropic method is not designed to tackle problems where
the \emph{only} information is data $D=(x_{1},x_{2}\ldots x_{n})$. It is not
that it gives a wrong answer; it gives no answer at all because there is no
constraint to impose; the entropic engine cannot even get started.

But there is a possible exception:\ surely the data $(x_{1},\ldots )$ must
be relevant to inferences about the quantity $x$ itself. More generally,
whether a given piece of data turns out to be relevant or not depends on
what is the question being asked. If we want to make inferences about a
particular function $f(x)$ then we know that information about $\langle
f(x)\rangle $ must surely be relevant --- in which case we deal with a type 
\textbf{B} constraint. Whether the information captured by $\langle
f(x)\rangle $ is sufficient for reliable inferences, or whether higher
moment constraints, $\langle f^{2}\rangle ,\langle f^{3}\rangle \ldots $
must also be included is a question that must be addressed on a case by case
basis.

\noindent \textbf{Example --- a type B constraint plus data:}

Suppose then, that in addition to the data $D=(x_{1},x_{2}\ldots x_{n})$
collected in $n$ independent experiments we have information described as
type \textbf{B} in the previous section: the expectation $\langle f\rangle $
captures relevant information. Then we can proceed to maximize the entropy $%
S[p,q]$, where $q(x_{i})$ is a (possibly uniform) prior distribution,
subject to the constraint $\langle f\rangle =F$ where the unknown $F$ is
treated as a free parameter. If the variable $x$ can take $k$ discrete
values labeled by $i$ we let $q(x_{i})=q_{i}$ and $f(x_{i})=f_{i}$ and the
result is a canonical distribution 
\begin{equation}
p(x_{i}|\lambda )=\frac{1}{Z}q_{i}e^{-\lambda f_{i}}~,  \label{canon dist}
\end{equation}%
where%
\begin{equation}
Z=\sum\limits_{i=1}^{k}q_{i}\,e^{-\lambda f_{i}}\quad \text{and}\quad
\langle f\rangle =-\frac{\partial \log Z}{\partial \lambda }
\label{Z and Ef}
\end{equation}%
with an unknown multiplier $\lambda $ that can be estimated from the data $D$
using standard Bayesian methods. Assuming the $n$ experiments are
independent then Bayes rule gives, 
\begin{equation}
p(\lambda |D)=\frac{p(\lambda )}{p(D)}\prod\limits_{j=1}^{n}\frac{1}{Z}%
q_{j}e^{-\lambda f_{j}}~,
\end{equation}%
where $p(\lambda )$ is the prior for $\lambda $. It is convenient to
consider the logarithm of the posterior, 
\begin{equation}
\log p(\lambda |D)=\log p(\lambda )-\log p(D)-\sum\limits_{j=1}^{n}(\log
Z-\log q_{j}+\lambda f_{j})~.  \notag
\end{equation}%
The value of $\lambda $ that maximizes the posterior $p(\lambda |D)$ is such
that 
\begin{equation}
0=\frac{\partial \log p(\lambda )}{\partial \lambda }-n\frac{\partial \log Z%
}{\partial \lambda }-n\bar{f}~.
\end{equation}%
where $\bar{f}$ is the sample average, 
\begin{equation}
\bar{f}=\frac{1}{n}\sum\limits_{j=1}^{n}f_{j}~.
\end{equation}%
Using (\ref{Z and Ef}) we see that the expected value $\langle f\rangle $
(and its corresponding $\lambda $) can be estimated from the data as 
\begin{equation}
\langle f\rangle =\bar{f}-\frac{1}{n}\frac{\partial \log p(\lambda )}{%
\partial \lambda }~.  \label{Ef}
\end{equation}%
As $n\rightarrow \infty $ the second term on the right hand side vanishes
and we see that the optimal $\lambda $ is such that $\langle f\rangle =\bar{f%
}$. This is to be expected: as is usual in Bayesian inference for large $n$
the data $D$ overwhelms the prior $p(\lambda )$ and $\bar{f}$ tends to $%
\langle f\rangle $ (in probability). But the result eq.(\ref{Ef}) also shows
that when $n$ is not large then the prior can make a non-negligible
contribution. In general one should not assume that $\langle f\rangle
\approx \bar{f}$ .

Let us emphasize that this analysis holds only when there is additional
knowledge to the effect that the specific variable $f$ captures relevant
information. In the absence of such knowledge we are back to the previous
example -- just data -- and we have no reason to prefer the function $f(x)$
over any other function $g(x)$ and accordingly we have no reason to prefer
the distribution (\ref{canon dist}) over any other canonical distribution $%
q_{i}e^{-\lambda g_{i}}/Z$.\footnote{%
In \cite{Jaynes 1979} (pp.72-75) Jaynes discussed the \textquotedblleft 
constraint rule\textquotedblright\ by carrying out a similar MaxEnt/maximum
likelihood analysis which failed to include the contribution from the prior
for $\lambda $ [the second term on the right of eq.(\ref{Ef})]. His
conclusion that we are always justified to set $\langle f\rangle =\bar{f}$
is therefore doubly wrong --- it is both necessary that $f$ reflect relevant
information and that the correction due to the prior be negligible.}

\section{Confusion about ignorance}

To set the stage for the issues involved consider a three-sided die. Its
faces are labeled by the number of spots $i=1,2,3$ and have probabilities $%
\theta _{1},\theta _{2},\theta _{3}$ which will be collectively denoted by $%
\theta $. The space of distributions is the simplex $\mathcal{S}_{2}$ with $%
\textstyle\sum\nolimits_{i}\theta _{i}=1$ as shown in the figure. A fair die is one
for which $\theta =\theta _{C}=(\frac{1}{3},\frac{1}{3},\frac{1}{3})$ which
lies at the very center of the simplex. The expected number of spots for a
fair die is $\langle i\rangle =2$. Having $\langle i\rangle =2$ is no
guarantee that the die is fair but if $\langle i\rangle \neq 2$ the die is
necessarily biased.

\begin{figure}
\includegraphics[
trim=1.5in 2.5in 3.0in 1.75in,
natheight=10.00in,
natwidth=10.00in,
height=2.40in, 
width=3.52in] {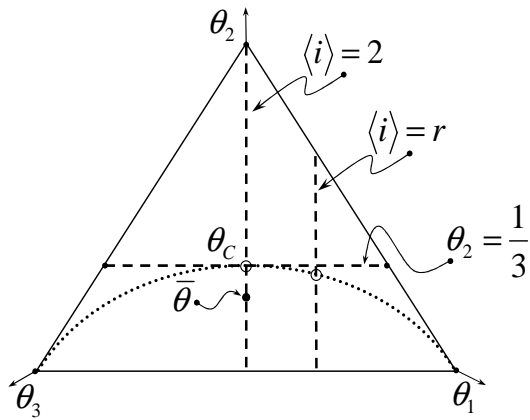}
\caption{Constraints $\langle i\rangle=r$ are shown as vertical lines on the
simplex. The dotted line is the set of MaxEnt distributions $\theta_{ME}(r)$
as $r$ spans the range from $1$ to $3$. If $r$ is unknown the average of
$\theta_{ME}(r)$ over $r$ leads to the distribution marked by $\bar{\theta}$.}\label{fig8-2}
\end{figure}

The paradox discussed by Friedman and Shimony \cite{Friedman Shimony 1971}
and further explored in \cite{Shimony 1985}, \cite{Seidenfeld 1986} and \cite%
{Uffink 1996} arises from analyzing a situation of complete ignorance in two
ways that are (mistakenly) thought to be equivalent.

Here is the first way to express complete ignorance:

\noindent \textbf{Ignorance}$_{1}$ \textbf{---} Nothing is known about the
die; we do not know that it is fair but on the other hand there is nothing
that induces us to favor one face over another. On the basis of this minimal
information we can use MaxEnt. Maximize 
\begin{equation}
S(\theta )=-\sum\nolimits_{i}\theta _{i}\log \theta _{i}  \label{Sx}
\end{equation}%
subject to $\sum\nolimits_{i}\theta _{i}=1$ and the resulting maximum
entropy distribution is $\theta _{ME}=\theta _{C}$.

To set the background for the second way to represent complete ignorance
consider constraints of the form $\langle i\rangle =r$ with $1\leq r\leq 3$.
In the figure the constraints corresponding to $r=2$ and to a generic $r$
are shown as vertical dashed lines. Maximizing $S(\theta )$ subject to $%
\langle i\rangle =2$ and normalization leads us to assign $\theta
_{ME}=\theta _{C}$ at the center of the simplex. Maximizing $S(\theta )$
subject to $\langle i\rangle =r$ and normalization leads to the point where
the $r$ line crosses the dotted line. The dotted curve is the set of MaxEnt
distributions $\theta _{ME}(r)$ as $r$ spans the range from $1$ to $3$.

Here is the second way to express complete ignorance:

\noindent \textbf{Ignorance}$_{2}$ \textbf{--- }It is tempting (but wrong!)
to pursue the following line of thought. We have a die but we do not know
much about it. We do know, however, that the quantity $\langle i\rangle $
must have some value, call it $r$, about which we are ignorant too. Now, the
most ignorant distribution given $r$ is the MaxEnt distribution $\theta
_{ME}(r)$. But $r$ is itself unknown so a more honest assignment for $\theta 
$ would be an average over $r$, 
\begin{equation}
\bar{\theta}=\int dr\,p(r)\theta _{ME}(r)~,  \label{thetabar}
\end{equation}%
where $p(r)$ reflects our uncertainty about $r$. It may, for example, make
sense to pick a uniform distribution over $r$ but the precise choice is not
important for our purposes. The point is that since the MaxEnt dotted curve
is concave the point $\bar{\theta}$ necessarily lies below $\theta _{C}$ so
that $\bar{\theta}_{2}<1/3$. And we have a paradox: we started assuming
complete ignorance and through a process that claims to express full
ignorance at every step we reach the conclusion that the die is biased
against $i=2$. Where is the mistake?

Another way to expose the problem is to force the issue and impose that the
two descriptions of complete ignorance be equivalent by fiat, 
\begin{equation}
\theta _{C}=\bar{\theta}~.  \tag{wrong}
\end{equation}%
Since the dotted line is concave this equality can only be achieved if $%
p(r)=\delta (r-2)$. And, again, we have a paradox: we started admitting
complete ignorance about the die and therefore also about the value of $r$
and we end with complete certainty that $r=2$. Where is the mistake?

Before we blame the entropic method of inference it is best to take a closer
look at \textbf{Ignorance}$_{2}$. One clue is symmetry. A situation of
complete ignorance ought to treat the outcomes $i=1,2,3$ symmetrically but
the end result is a distribution that is biased against $i=2$. The symmetry
must have been broken somewhere and it is clear that this happened at the
moment we imposed the constraint on $\langle i\rangle =r$ which is shown as 
\emph{vertical} lines on the simplex. Had we chosen to express our ignorance
not in terms of the unknown value of $\langle i\rangle =r$ but in terms of
some other function $\langle f(i)\rangle =s$ then we could have easily
broken the symmetry in some other direction. For example, let $f(i)$ be a
cyclic permutation of $i$, 
\begin{equation}
f(1)=2,\quad f(2)=3,\quad \text{and}\quad f(3)=1~,
\end{equation}%
then repeating the analysis above would lead us to conclude that $\bar{\theta%
}_{1}<1/3$, which represents a die biased against $i=1$. Thus, the question
becomes: What leads us to choose a constraint on $\langle i\rangle $ rather
than a constraint on $\langle f(i)\rangle $ when we are equally ignorant
about both?

The discussion in section 2 is relevant here. \noindent The paradox with the
three-sided die arises because a constraint of type \textbf{D} has been
treated as if it were a constraint of type \textbf{B}. The correct approach
is to recognize that we do not know whether it is the constraint $\langle
i\rangle $ or any other function $\langle f\rangle $ that captures relevant
information and their numerical values $r$ are also unknown --- clearly a
type \textbf{D} constraint. There is nothing to single out $\langle i\rangle 
$ or any other $\langle f\rangle $ and therefore the correct inference
consists of maximizing $S$ imposing the only constraint \emph{we actually
know}, namely, normalization. The result is as it should be --- a uniform
distribution ($\theta _{ME}=\theta _{C}$) which agrees with \textbf{Ignorance%
}$_{1}$.

On the other hand, the \textbf{Ignorance}$_{2}$\textbf{\ }argument that led
to the assignment of $\bar{\theta}$ in eq.(\ref{thetabar}) and to $\bar{%
\theta}_{2}<1/3$ would have been correct if we actually had knowledge that
it is the particular variable $\langle i\rangle $ -- and not any other $%
\langle f\rangle $ -- that captures information that is relevant to this
very particular die. Thus, imposing the type \textbf{B} constraint $\langle
i\rangle =r$ when $r$ is unknown and then averaging over $r$ represents a
situation in which \emph{we know something}. There is some ignorance here --
we do not know $r$ -- but this is not extreme ignorance.

We can summarize as follows: knowing that the die is biased against $i=2$
but not knowing by how much (\textbf{Ignorance}$_{2}$) is not the same as
not knowing anything about the die (\textbf{Ignorance}$_{1}$).

A thermodynamic version of the paradox is discussed in \cite{Shimony 1985}.
Here is the background: A physical system can be in any of $n$ microstates
labeled $i=1\ldots n$. When we know absolutely nothing about the system (%
\textbf{Ignorance}$_{1}$) maximizing entropy subject to the single
constraint of normalization leads to a uniform probability distribution, 
\begin{equation}
p_{U}(i)=1/n~.
\end{equation}

A different (and wrong!) way to express complete ignorance (\textbf{Ignorance%
}$_{2}$) is to argue that the expected energy $\langle \varepsilon \rangle $
must have some value $E$ about which we are ignorant. Maximizing entropy
subject to both $\langle \varepsilon \rangle =E$ and normalization leads to
the usual Boltzmann distributions, 
\begin{equation}
p(i|\beta )=\frac{e^{-\beta \varepsilon _{i}}}{Z(\beta )}\quad \text{where}%
\quad Z(\beta )=\sum\limits_{i}e^{-\beta \varepsilon _{i}}~.
\end{equation}%
Since the inverse temperature $\beta =\beta (E)$ is itself unknown we must
average over $\beta $, 
\begin{equation}
\bar{p}(i)=\int d\beta \,p(\beta )p(i|\beta )~.
\end{equation}%
To the extent that both distributions are thought to reflect complete
ignorance we must impose 
\begin{equation}
p_{U}(i)=\bar{p}(i)  \tag{wrong again}
\end{equation}%
which can be shown (see \cite{Shimony 1985}) to imply that%
\begin{equation}
p(\beta )=\delta (\beta )\quad \text{or}\quad \beta =0~.
\end{equation}%
Indeed, setting the Lagrange multiplier $\beta =0$ in $p(i|\beta )$ leads to
the uniform distribution $p_{U}(i)$. And now we have a paradox: complete
ignorance about the system (\textbf{Ignorance}$_{1}$) implies we are
ignorant about its temperature. In fact, the system might not be in thermal
equilibrium in which case it may not even have a temperature at all. But we
also have the second way of expressing ignorance (\textbf{Ignorance}$_{2}$)
and if we impose that the two agree we are led to conclude that $\beta $ has
the value $\beta =0$ so that the temperature is precisely known. From
complete ignorance we have (wrongly) concluded that the system must be
infinitely hot --- confusion about ignorance is hell.

The paradox is dissolved once we realize that, just as with the die problem,
a type \textbf{D} constraint has been treated as type \textbf{B}. Knowing
nothing about a system means we do not know whether it is in equilibrium or
not. We do not know whether it is isolated and whether its energy is
conserved and therefore it is not clear that $\langle \varepsilon \rangle $
might even be relevant information. This is the kind of non-information we
earlier called a type \textbf{D} constraint that ought to be ignored.

On the other hand if we were to have actual knowledge that the system is in
thermal equilibrium then it would be legitimate to impose a constraint on
the expected energy $\langle \varepsilon \rangle $ --- as discussed in \cite%
{Caticha 2012} thermal equilibrium is the physical condition that singles
out the expected energy $\langle \varepsilon \rangle $ as being the relevant
piece of information. Since the temperature is unknown this is a type 
\textbf{B} constraint.

To summarize: knowing that a system is in thermal equilibrium while being
ignorant about its temperature is not the same as knowing nothing about the
system. 

It may be worthwhile to rephrase this important point in yet another way.
Let $i\in \mathcal{I}$ and $\beta \in \mathcal{B}$ where $\mathcal{I}$ is
the space of microstates and $\mathcal{B}$ is the space of some arbitrary
quantities $\beta $. The rules of probability theory allow us to write 
\begin{equation}
p(i)=\int d\beta \,p(i,\beta )\quad \text{where}\quad p(i,\beta )=\,p(\beta
)p(i|\beta )~.
\end{equation}%
Paradoxes will easily arise if we fail to distinguish a situation of
complete ignorance from a situation where we have actual knowledge about the
conditional probability $p(i|\beta )$, which is what gives the parameter $%
\beta $ its meaning as inverse temperature.

\section{Summary}

In entropic inference information is represented by constraints. In this
tutorial we have argued that when constraints are expressed in terms of expected
values the same formal expression $\langle f\rangle =F$ can be used to
represent four different types of available information and failure to
distinguish between them can lead to errors. 

In decreasing order of useful information the types range from type \textbf{A%
}, in which we know that the quantity $f$ is relevant to the inference at
hand and its expected value $F$ is known; to type \textbf{B}, in which $f$
is known to be relevant but $F$ is unknown; and type \textbf{C}, in which $f$
is of interest only because $F$ happens to be known. Constraints of type 
\textbf{A} and \textbf{B} are common and therefore important; type \textbf{C}
constraints are less so because information about expected values (as
opposed to sample averages) is less directly available.

Finally, there are situations of no information at all, labeled as type 
\textbf{D}, in which we know neither that $f$ is relevant nor its expected
value $F$. The point of identifying and labeling the non-informative type 
\textbf{D} is precisely as to avoid the errors that arise from confusing 
\textbf{D} with the more informative types \textbf{A}, \textbf{B}, and 
\textbf{C}. 

The usefulness of distinguishing these four types was illustrated by
discussing the constraint rule and the Friedman-Shimony paradox. 
\newline
\noindent \textbf{Acknowledgements: }I am grateful to N\'{e}stor Caticha, R.
Fischer, A. Giffin, A. Golan, R. Preuss, C. Rodr\'{\i}guez, T. Seidenfeld,
J. Skilling and J. Uffink for many useful discussions on entropic inference.

\end{document}